\documentclass[aps,prl,twocolumn,showpacs]{revtex4}
\usepackage{amsmath,bm}
\usepackage{graphicx,epsfig,braket,amssymb,amsfonts}
\DeclareGraphicsExtensions{.pdf,.png,.jpg}

\renewcommand{\v} [1]{{\bf #1}}
\newcommand{\ba}{\begin{eqnarray}}
\newcommand{\ea}{\end{eqnarray}}
\newcommand{\nn}{\nonumber \\}
\newcommand{\bpm}{\begin{pmatrix}}
\newcommand{\epm}{\end{pmatrix}}

\begin{document}

\title{Nature of Orbital and Spin Rashba Coupling in the \\Surface Bands of SrTiO$_3$ and KTaO$_3$}

\author{Panjin Kim}
\affiliation{Department of Physics, Sungkyunkwan University, Suwon 440-746, Korea}
\author{Kyeong Tae Kang}
\affiliation{Department of Physics, Sungkyunkwan University, Suwon 440-746, Korea}
\author{Jung Hoon Han}
\email[Electronic address:$~~$]{hanjh@skku.edu}
\affiliation{Department of Physics, Sungkyunkwan University, Suwon 440-746, Korea}

\begin{abstract}
Tight-binding models for the recently observed surface electronic bands of SrTiO$_3$ and KTaO$_3$ are analyzed with a view to bringing out the relevance of momentum-space chiral angular momentum structures of both orbital and spin origins.
Depending on the strength of electric field associated with inversion symmetry breaking at the surface, the orbital and the accompanying spin angular momentum structures reveal complex linear and cubic dependencies in the momentum $\mathbf{k}$ (linear and cubic Rashba effects, respectively) in a band-specific manner.
Analytical expressions for the cubic orbital and spin Rashba effects are derived by way of unitary transformation technique we developed, and compared to numerical calculations.
Due to the $C_{4v}$ symmetry of the perovskite structure the cubic Rashba effect appears as in-plane modulations.
\end{abstract}
\maketitle

The discovery of surface electronic states in strontium titanate (SrTiO$_3$)\cite{syro-nature,shen-NM} has stirred great excitement at the time the material is being viewed as a critical component of the emerging field of oxide electronics~\cite{oxide-review}.
The origin of surface states in SrTiO$_3$ and a related material KTaO$_3$~\cite{KTO-shen,KTOprb} (STO and KTO for short, respectively) is currently under active investigation~\cite{KTOprb,held}.
Both materials' surface states originate from $t_{2g}$-orbitals whose relevant tight-binding parameters for the electronic structure are largely determined, including the one pertaining to the degree of inversion symmetry breaking (ISB) at the surface~\cite{KTOprb,held}.

Several features make STO and KTO surface states an ideal ground for the study of Rashba-related phenomena.
First is the way that Rashba effects would play out among the several observed bands of differing orbital characters.
ARPES measurements up to now~\cite{syro-nature,shen-NM,KTO-shen,KTOprb} did not clearly resolve the Rashba-split bands, presumably due to the smallness of the predicted Rashba parameter~\cite{king-texture}.
Transport measurements do reveal the Rashba term, of cubic order in momentum, through analysis of the orientation-dependent magneto-resistance data on STO surface~\cite{kimura,kimura-KTO}.
Existing theories treat Rashba effects of $t_{2g}$-derived bands phenomenologically~\cite{held,nayak} and cannot, for instance, explain the complex band-specific spin and orbital angular momentum structures observed in the electronic structure calculation~\cite{king-texture}.

It has recently been argued that multi-orbital bands, subject to the surface ISB electric field, must give rise to an entity called the chiral orbital angular momentum (OAM) in momentum space~\cite{chiral-OAM,kim-OAM}.
The argument remains valid as long as the crystal field splitting does not quench the multi-orbital degrees of freedom in a given band structure.
Such conditions seem to be well met in both STO and KTO, leading to the term $\sim\v k \times \v E \cdot \v L$ where $\v L$ is the OAM operator for $t_{2g}$-orbitals, $\mathbf{k}$ is the linear momentum, and $\mathbf{E}$ is the surface-normal electric field.
The effect was dubbed ``orbital Rashba effect"~\cite{chiral-OAM} in analogy to the similar chiral structure of spins on the surface~\cite{rashba}.
It was shown that pre-existing chiral OAM structure implies the linear Rashba effect upon the inclusion of spin-orbit interaction~\cite{chiral-OAM}.

The ideas and techniques developed in Ref.~\onlinecite{chiral-OAM} can be extended to address cubic-order Rashba effects, of both orbital and spin origins as we show now. The complex interplay of spin and orbital textures in momentum space can be understood in a systematic and band-specific manner through such analysis.
Based on $t_{2g}$-orbital models pertinent to STO and KTO surfaces, we derive the cubic orbital and spin Rashba terms consistent with the $C_{4v}$-symmetry of the perovskite structure.
Earlier derivation of cubic Rashba term for the $C_{3v}$-symmetric single-band surface of topological insulator predicted the coupling of cubic momentum to the out-of-plane spin component~\cite{warping}.
The $C_{4v}$-symmetry of perovskite surface in contrast requires cubic momentum dependence in the in-plane modulation of orbital and spin textures.

The tight-binding Hamiltonian we employ is the same as already discussed in several papers~\cite{KTOprb,held,chiral-OAM-in-magnetic-band,anomalous-hall}.

\begin{align}\label{h1r}
&{\cal H} = \sum_{\v k, \sigma} \v C^\dag_{\v k, \sigma} {\cal H}_0 \v C_{\v k, \sigma} + \sum_{\v k} \v C^\dag_{\v k } {\cal H}_\mathrm{so} \v C_{\v k} , \nn
&{\cal H}_0 = - \left(\begin{array}{ccc}
    2t(c_x \!+\!c_y)\!-\!\delta & -2i\gamma s_x  & -2i\gamma s_y \\
    2i\gamma s_x & 2tc_y\!+\!2t'c_x & 0 \\
    2i\gamma s_y  & 0 & 2tc_x\!+\!2t'c_y\\
  \end{array}\right) , \nn
&{\cal H}_{\rm so} = \lambda_{\rm so} \v L \cdot \bm \sigma .
\end{align}
Spin-orbit matrix ${\cal H}_{\rm so}$ in its explicit form can be found in Refs. \cite{held,chiral-OAM-in-magnetic-band}.
The second-quantized operators are written in the basis $(\sigma=\uparrow,\downarrow)$

\begin{align} & \v C_{\v k,\sigma}=(c_{\v k,xy,\sigma},  c_{\v k,yz,\sigma}
, c_{\v k,zx,\sigma})^\intercal, \nn
& \v C_{\v k} = (c_{\v k,xy,\uparrow}, c_{\v k,xy,\downarrow}, c_{\v
k,yz,\uparrow}, c_{\v k,yz,\downarrow}, c_{\v k,zx,\uparrow}, c_{\v
k,zx,\downarrow})^\intercal . \nonumber
\end{align}
Furthermore, $t$ and $t'$ are $\sigma$- and $\pi$-bonding parameters of $t_{2g}$-orbitals, respectively.
Abbreviations for the momentum dependence are $c_{x(y)}=\cos k_{x(y)}$, $s_{x(y)}=\sin k_{x(y)}$.
The lattice constant is taken to be unity.
Terms proportional to $\gamma$ in ${\cal H}_0$ mediate parity-violating hopping processes that can be cast as $\hat{z}\cdot(\mathbf{k} \times \mathbf{L})$ around the $\Gamma$ point~\cite{chiral-OAM}.
The energy difference between $xy$ and $yz~(zx)$ orbitals is summed up as $\delta$ accounting for different responses of in-plane vs. out-of-plane orbitals to surface confinement, surface crystal field, as well as lack of hopping terms in the $z$-direction compared with the bulk system~\cite{KTOprb,held}.
Numerical estimates of all tight-binding parameters extracted from Refs. \onlinecite{KTOprb,held} are summarized in Table \ref{parameters}.
Despite KTO having a much larger spin-orbit energy $\lambda_{\rm so}$ than STO, the overall surface band structures are qualitatively similar due to the even larger bandwidths set by $t$ and $t'$.

\begin{table}[htbp]
 \centering
  \begin{tabular}{|c||c| c| c| c| c| }
  \hline
  unit (eV)  &    $t$  & $t'$      & $\delta$  &$\lambda_{\rm{so}} $ &$\gamma$  \tabularnewline\hline
  SrTiO$_3$  &~~0.277~~& ~~0.031~~ & ~~0.092~~ &~~0.01~~             &~~0.02~~  \tabularnewline
  KTaO$_3$   &~~0.75~~ & ~~0.075~~ & ~~1.0~~   &~~0.15~~             &~~0.01~~  \tabularnewline
  \hline
  \end{tabular}
\caption{Tight-binding parameters for STO and KTO. STO parameters are taken from Ref. \cite{held}.
KTO parameters are obtained from best fits to the graphs shown in Ref. \cite{KTOprb}. } \label{parameters} \end{table}

The ISB part of ${\cal H}_0$, being proportional to $\mathbf{k}$, diminishes as the $\Gamma$ point is approached, where also the energy gap $E_{\rm gap}$ between $d_{xy}$-orbital-derived lowest energy band and the other two bands exist ($E_{\rm gap}=2\Delta \approx 0.4 ~(0.35)$ eV for STO (KTO), and $\Delta=t-t'-\delta/2$).
Such circumstances invite a unitary rotation ${\cal U}_1$~\cite{chiral-OAM}, which schematically renders a new Hamiltonian

\begin{align}\label{U1V1}
{\cal U}_1^\dag{\cal H}_0 {\cal U}_1 = \left(\begin{array}{ccc}
    {\cal H}_{11}       & {\cal H}_{12}(k^3)  & {\cal H}_{13}(k^3) \\
    {\cal H}_{21}(k^3)  & {\cal H}_{22}       & 0                  \\
    {\cal H}_{31}(k^3)  & 0                   & {\cal H}_{33}      \\
  \end{array}\right).
\end{align}
Detailed derivations are included in the Supplementary Information (SI)~\cite{SI}.
The unitary-transformed Hamiltonian resembles the old one ${\cal H}_0$ in form, except the off-diagonal elements coupling the lowest-energy band to the rest now appear at the {\it cubic order or higher}.
By comparison the bare Hamiltonian ${\cal H}_0$ had linear-$k$ terms as off-diagonal elements.
The observation prompts us to seek a further unitary transformation that might render even higher-order off-diagonal terms.
Indeed it is possible to carry out a second unitary rotation ${\cal U}_2$ that produces off-diagonal elements between the lowest-energy band and the remaining bands of {\it fifth order} in momentum.
For considerations of at most cubic order in momentum such terms can now be discarded.

We emphasize that the current technique differs vastly in spirit from the early work of Winkler, who derived cubic Rashba terms specific to the strongly spin-orbit-coupled $J=3/2$ and $J=1/2$ bands~\cite{winkler}.
Spin-orbit interaction (SOI) is not yet included at this stage, hence the results obtained will pertain to linear and cubic modulations of the \textit{orbital angular momentum}.

OAM averages of each band are given by the expectation values of $\v L = (\sum_i \v L_i )/N$, for the atomic OAM operator $\v L_i$ in the $t_{2g}$-orbital space at site $i$, and $N$ denoting the number of atomic sites (for details see Ref.~\onlinecite{chiral-OAM}).
Based on the diagonalization scheme just outlined, we can work out the OAM averages in two limiting cases: $\gamma \gg \Delta$ and $\gamma \ll \Delta$.
A more complete expression of the OAM average valid for arbitrary ratios of $\gamma/\Delta$ can be found in the SI.
The first, $\gamma \gg \Delta$ limit may be formal, since $\gamma$ arising from the surface confinement energy is usually smaller than the energy gap $\Delta$.
Still, this is a useful limit to consider as each band will be assigned a clear sense of orbital chirality.
For $\gamma \gg \Delta$ the OAM averages become

\begin{align}\label{largegamma}
&L_+^{(1)} \approx -2 i \frac{\gamma}{\Delta} k_+ + i \frac{\gamma}{\Delta}\left[\frac{1}{12}-\frac{t-t'}{4\Delta}\right] k_-^3,\nn
&L_+^{(-1)}\approx 2 i \frac{\gamma}{\Delta} k_+ - i \frac{\gamma}{\Delta}\left[\frac{1}{12}-\frac{t-t'}{8\Delta}\right] k_-^3,\nn
&L_+^{(0)} \approx i \frac{\gamma}{\Delta} \frac{t-t'}{8\Delta} k_-^3,
\end{align}
$k_\pm = k_x \pm i k_y$. $L_+^{(m)}$ denotes $\langle \psi_{\v k}^{(m)} | (L_x \! + \! i L_y ) |\psi_{\v k}^{(m)} \rangle$, an average for the quasi-exact eigenstate $|\psi_{\v k}^{(m)} \rangle$ at momentum $\v k$, labeled with the energy band index $m$.
It is found for arbitrary ratio $\gamma/\Delta$ that $m=1$ band stays at the lowest energy $E^{(1)}$ whereas bands with $m=-1$ and $m=0$ have lighter and heavier effective masses with the energies $E^{(-1)}$ and $E^{(0)}$, respectively.
In addition to referring to the energy band, $m$ reflects the orbital chirality for each band in the $\gamma \gg \Delta$ limit.
For example, $m=1~(-1)$ band has clockwise (counterclockwise) rotation of OAM averages in the momentum space and $m=0$ band does not have finite OAM average up to the first order in $\v k$.
$L_+^{(1)}$ and $L_+^{(-1)}$ show coexistence of linear and cubic in-plane modulations which eventually drive the same effects on spin via SOI.

\begin{figure}[tp]
\includegraphics[width=0.48\textwidth]{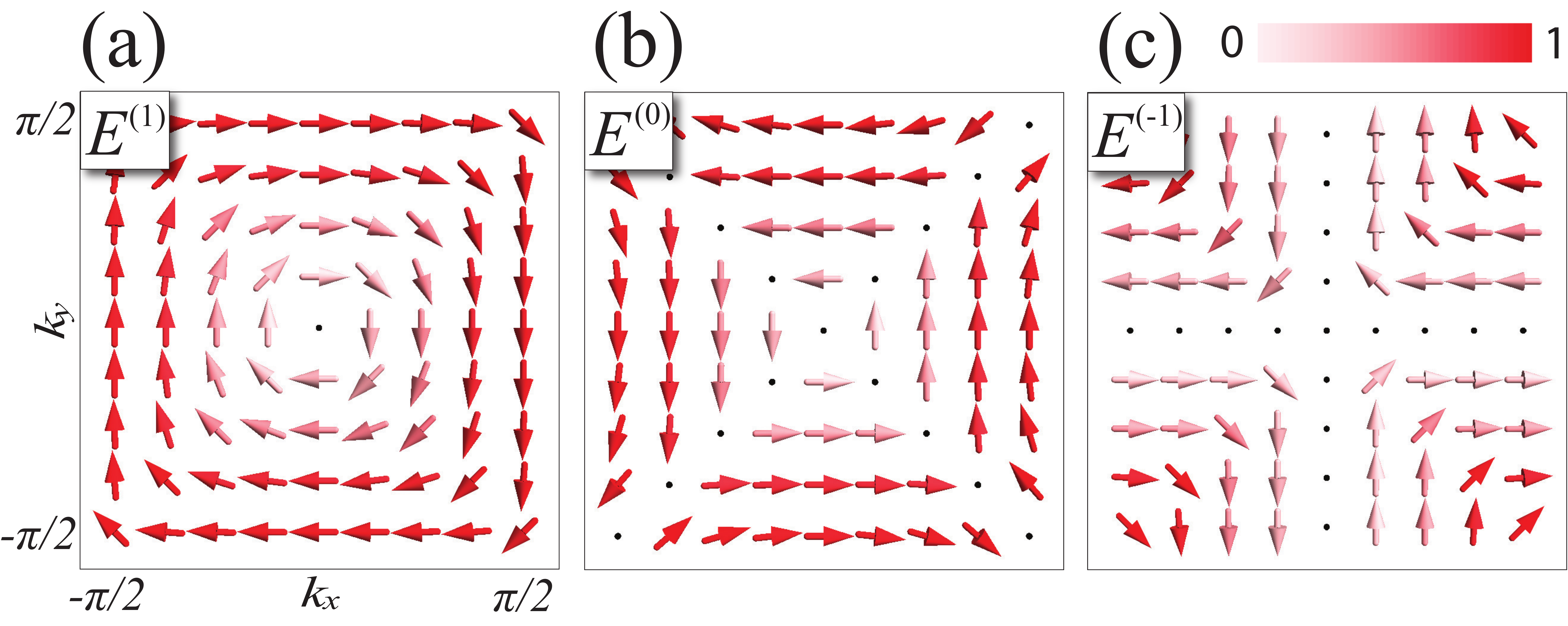}
\caption{(color online) (a)-(c) OAM averages in the $\gamma \ll \Delta$ limit for $m=1,0,-1$ bands obtained from numerical diagonalization of Eq. (\ref{h1r}).
Parameters are for STO (Table \ref{parameters}) with $\lambda_{\rm so} = 0$.
Magnitude of OAM is represented on a false-color scale in units of $\hbar$.
Directions and magnitudes of OAM vectors are consistent with Eq. (\ref{smallgamma}).
} \label{fig:OAM-STO}
\end{figure}

The situation is more complicated in the other limit $\gamma\ll \Delta$ which is relevant to the surface of STO and KTO.
OAM averages take the forms

\begin{align}\label{smallgamma}
&L_+^{(-1)} ~(L_+^{(0)}) \approx \left\{
                      \begin{array}{ll}
                        P_x ~(P_y), & \hbox{$|k_x|>|k_y|$,} \\
                        P_y ~(P_x), & \hbox{$|k_x|<|k_y|$,} \\
                        P_x+P_y ~(0), & \hbox{$|k_x|=|k_y|$,} \\
                      \end{array}
                    \right. \nn
& P_x = i \frac{\gamma}{\Delta} (k_+ \! - \! k_-) + \frac{i}{2} \frac{\gamma}{\Delta} \left[\frac{1}{12}\!-\!\frac{t-t'}{4\Delta}\right](k_+^3 \!-\! k_-^3 ) ,\nn
& P_y = i \frac{\gamma}{\Delta} (k_+ \! + \! k_-) - \frac{i}{2} \frac{\gamma}{\Delta} \left[\frac{1}{12}\!-\!\frac{t-t'}{4\Delta}\right](k_+^3 \!+\! k_-^3 ) .
\end{align}
%
Here $P_x$ and $P_y$ indicate the linearly polarized OAM in $k_x$- and $k_y$-directions, respectively.
$L_+^{(1)}$ retains the same form as in Eq. (\ref{largegamma}), conserving a definite sense of chirality for all ratios $\gamma /\Delta$.
On the other hand, two adjacent $m=0$ and $m=-1$ bands lose their sense of chirality for small ISB parameter.
Instead, each band has linearly polarized OAM that changes direction abruptly at every $90^{\circ}$ rotation of the momentum $\v k$, as shown in Fig. \ref{fig:OAM-STO} and predicted by Eq. (\ref{smallgamma}).
Such quasi-linear OAM patterns for the upper two bands of STO surface states is a strong, testable prediction that can be confirmed readily by the recently popular technique of circular dichroism ARPES~\cite{kim-OAM,chiral-OAM,OAM-Cu-Au}.

Next step is to examine the effect of atomic spin-orbit interaction ${\cal H}_\mathrm{so}$ on the band structure.
Although its influence is being considered {\it after} the ISB effects, its strength need not be smaller.
The philosophy behind the procedures up to this point was to \emph{find the right set of basis in the presence of ISB}, accurate up to third order in $\mathbf{k}$.
Once that basis has been identified, further interactions such as ${\cal H}_\mathrm{so}$ can be written down within the new basis.
Spin-orbit Hamiltonian $H_{\rm so} = \lambda_{\rm so} \bm \sigma \cdot \v L$ expressed in the unitary-transformed basis has matrix elements

\begin{align}\label{Hso-elements}
\langle m,\sigma |{\cal  H}_{\rm so} | m', \sigma' \rangle = \lambda_{\rm so} (\bm \sigma)_{\sigma, \sigma'}  \cdot ({\cal U}^\dag \v L {\cal U} )_{m, m' } , \end{align}
where ${\cal U}={\cal U}_1{\cal U}_2$. It means, in particular, that each $2\!\times\! 2$ diagonal block of the spin-orbit matrix pertaining to the band index $m$, abbreviated ${\cal H}_{\rm so}^{(m)}$, is precisely

\begin{align}\label{Hsoc}
 {\cal H}^{(m)}_{\rm so} = \lambda_{\rm so} \big[ \sigma_+ L_-^{(m)} + \sigma_- L_+^{(m)} \big]  ,
\end{align}
$\sigma^\pm =(\sigma^x \pm i \sigma^y )/2$, and $L^{(m)} _- = (L^{(m)} _+ )^*$.
Barring inter-band effects, it implies that \emph{spin Rashba splitting is a direct image of the corresponding momentum-space polarization of orbitals}.

\begin{figure}[hbp]
\includegraphics[width=0.48\textwidth]{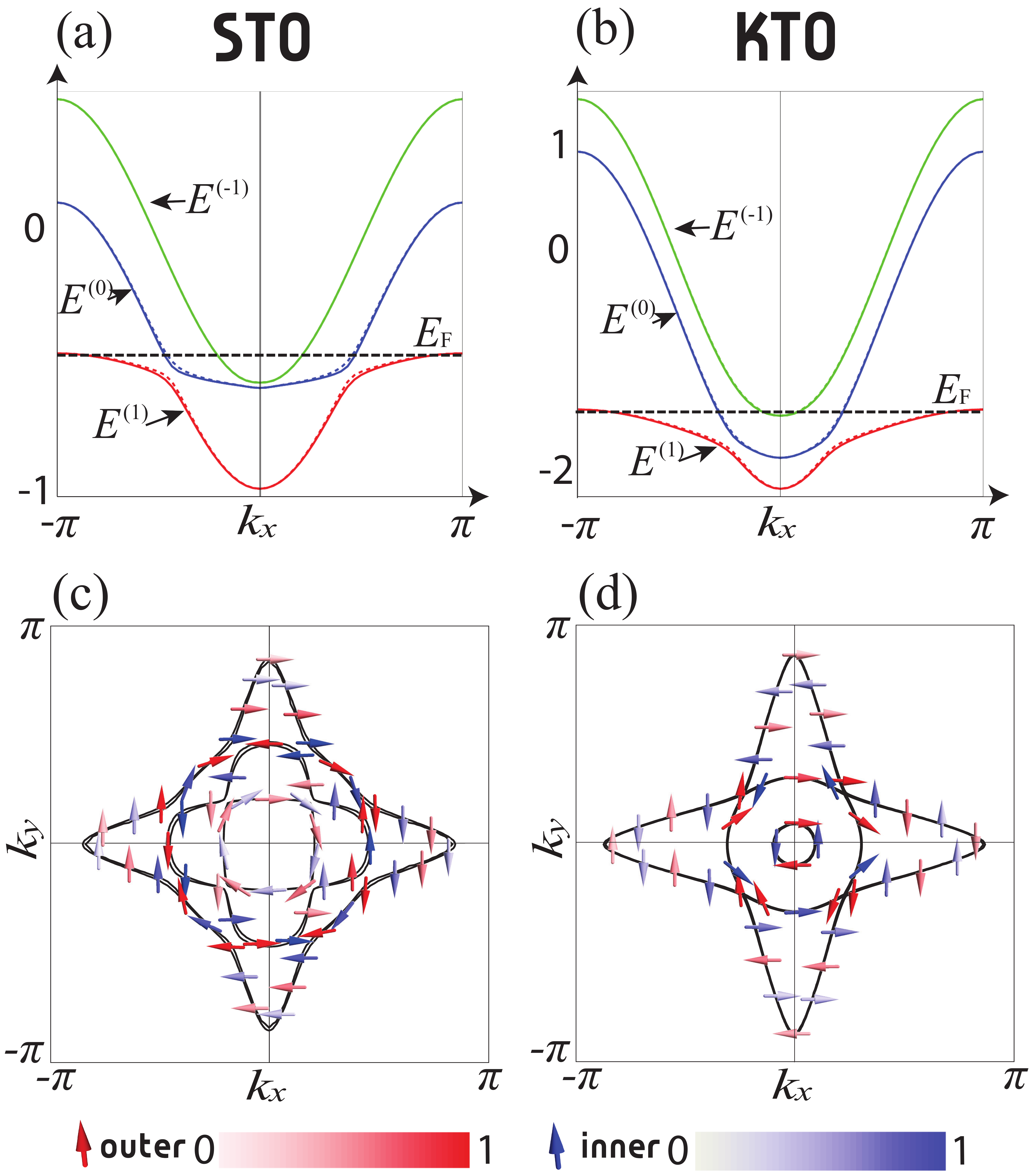}
\caption{(color online)
 (a),(b) $t_{2g}$-band structures in STO and KTO obtained from respective tight-binding models. For each subband pair, outer (inner) band is denoted as solid (dotted) line.
 Energy on the $y$-axis is in units of eV.
 (c),(d) Fermi surface at the energy $E_{\rm F}$ indicated in (a),(b) and corresponding OAM average vectors in STO and KTO.
 OAM averages of outer (inner) bands are represented as red (blue) arrows imposed on the Fermi surface contour.
 Magnitude of OAM is represented on a false-color scale in units of $\hbar$.
 Directions of OAM vectors in outer and inner Rashba-split states are the same in STO~\cite{king-texture}, whereas those in KTO are opposite.
} \label{fig:band-FS}
\end{figure}

\begin{figure*}[htbp]
\includegraphics[width=0.99\textwidth]{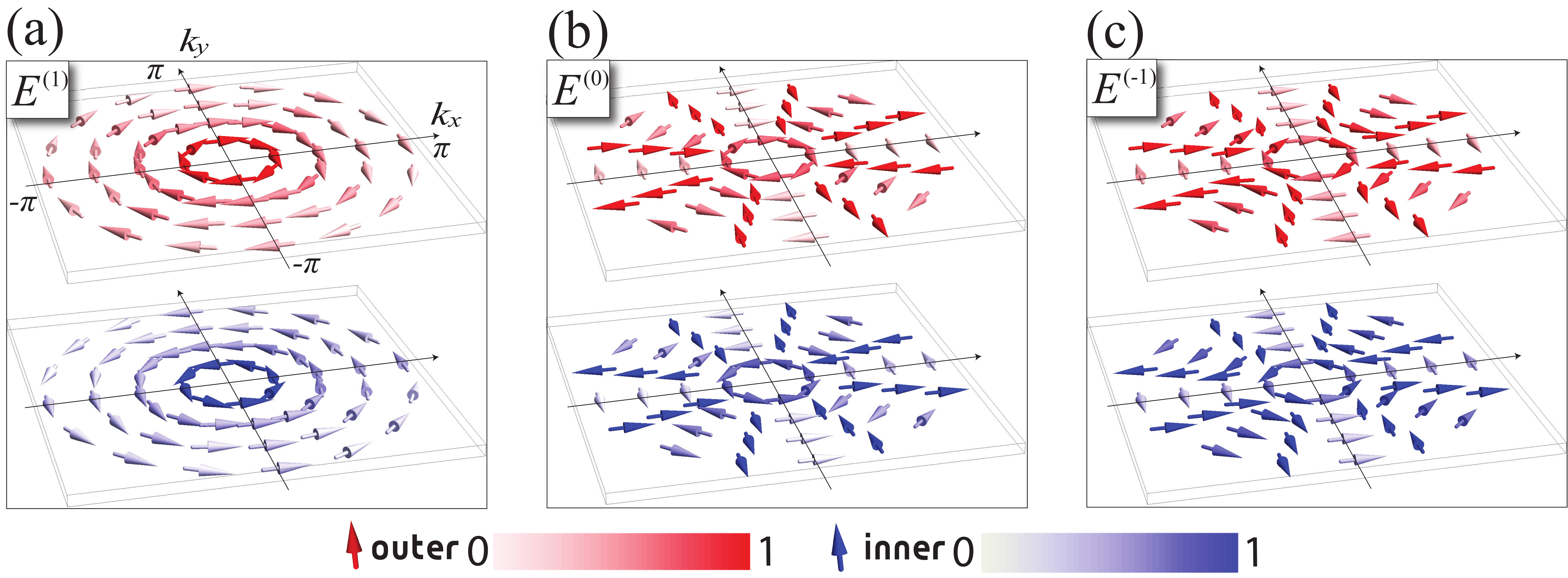}
\caption{(color online)
 (a)-(c) Band-specific OAM vectors in KTO obtained from the first principles calculation.
 In each figure, upper (lower) plot depicts the OAM structure of outer (inner) band.
 Each subband pair reveals opposite OAM direction for outer and inner bands.
 In addition to the rotating pattern around the $\Gamma$ point in all the bands, significant amount of linear-polarized OAM vectors are present in the $E^{(0)}$ and $E^{(-1)}$ bands.
} \label{fig:KTO}
\end{figure*}

Symmetry consideration provides the intuitive basis to understand the nature of linear and cubic Rashba coupling. Inter-band effects are negligible for the lowest-lying band separated from others by the gap $2\Delta$, thus permitting a reliable effective $2\!\times\!2$ Hamiltonian description.
Such Hamiltonian $H_{\rm eff}^{(1)}$ constructed by the theory of invariants must preserve all the symmetries of the physical system.
$C_{4v}$ symmetry is specified by two- and four-fold rotations $C_2,~C_4$, and mirror operations $M_v: (x,y) \rightarrow (-x,y)$ and $M_d: (x,y) \rightarrow (y,x)$.
Satisfying invariant conditions, the Rashba Hamiltonian can only take the following form up to the third order in momentum $\mathbf{k}$,

\begin{align}\label{Heff}
  H_{\rm eff}^{(1)} \approx \alpha_{\rm R}^l (k_x \sigma_y \!-\! k_y \sigma_x) + i\alpha_{\rm R}^c (k_+^3 \sigma_+ \!-\! k_-^3 \sigma_-).
\end{align}
Here the first and the second terms reflect the linear and cubic Rashba effect, respectively.
Equation (\ref{Hsoc}) for $m=1$ takes precisely this form dictated by symmetry, with parameters calculated from the more microscopic approach.
There is a clear difference between $C_{3v}$- and $C_{4v}$-symmetric systems regarding the nature of cubic Rashba coupling.
While on $C_{3v}$-symmetric surface of topological insulators cubic terms in momentum $\mathbf{k}$ take the form $\sim (k_+^3 + k_-^3)\sigma_z$ implying the out-of-plane spin polarization~\cite{warping}, on $C_{4v}$-symmetric perovskite surface corresponding term reads $\sim (k_+^3 \sigma_+ + k_-^3 \sigma_-)$, leading to the in-plane spin structures.

Unlike the $m=1$ band, the $m=0$ and $m=-1$ bands touch at $\v k=0$ with a non-negligible inter-band matrix elements due to the spin-orbit effect as shown in Eq. (\ref{Hso-elements}).
Figure~\ref{fig:band-FS} describes the band structures and the OAM average vectors on the Fermi surfaces in STO and KTO~\cite{king-texture,KTOprb} obtained from the full numerical diagonalization of Eq.~(\ref{h1r}).
Although the band structures of STO and KTO surface states are qualitatively similar due to the common $t_{2g}$-orbital and crystal structures, the order-of-magnitude difference in their respective spin-orbit coupling strengths clearly manifest themselves in the spin and orbital angular momentum textures.
In Fig.~\ref{fig:band-FS}(c), directions and magnitudes of OAM vectors are nearly identical in each of the outer and inner spin-split subband pair, indicating the OAM features predicted in Eq.~(\ref{smallgamma}) survive in STO even under the influence of SOI.
Spin angular momentum, on the other hand, is fully polarized in opposite directions in each subband pair, either parallel or anti-parallel to the underlying OAM (see SI for explicit spin structures).
KTO, on the contrary, has outer and inner Rashba-split states with opposite relative orientations of both orbital and spin vectors, as a consequence of the much stronger SOI strength.
Focusing on the Rashba effect, the maximum Rashba spin splitting along the $\Gamma - X$ direction occurs around $|\mathbf{k}| \sim 0.44\pi~(0.25\pi)$ with splitting energy $\Delta E_{\rm spin} \sim 20{\rm meV}~(28 {\rm meV})$ in STO (KTO).
Note that despite the order-of-magnitude difference in SOI strength, the Rashba spin splittings in STO and KTO show comparable energy scale.
Orbital and spin angular momentum structures obtained from density functional theory calculations~\cite{DFT} show the agreement with the full numerical diagonalization of Eq.~(\ref{h1r}).
Figure \ref{fig:KTO} illustrates the band-specific OAM vectors of KTO based on the first principles calculation.
Each subband pair has opposite OAM directions for outer and inner bands in entire Brillouin zone.

Since the expected Rashba spin splitting is very small, there appears slim chance of its unambiguous detection with the current resolution of the ARPES setup.
The chiral structure of OAM, on the other hand, should be readily observable by means of circular dichroism ARPES~\cite{kim-OAM,OAM-Cu-Au}.
The two would-be Rashba-split bands possess the \emph{same orbital chirality} in STO, and therefore give rise to the same circular dichroism response.
For KTO where the relative OAM directions cancel within the Rashba band pair, some other means of detecting their presence must be devised.

\acknowledgments{This work is supported by the NRF grant (No. 2013R1A2A1A01006430). P. K. acknowledges support from the Global Ph. D. Fellowship Program (NRF-2012). J. H. H. is indebted to Andreas Santander-Syro for discussion of his experiments during the APCTP workshop ``Bad Metal Behavior and Mott Quantum Criticality" in 2013.}

\end{document}


\title{Nature of Orbital and Spin Rashba Coupling in the \\Surface Bands of SrTiO$_3$ and KTaO$_3$: Supplementary Information}

\author{Panjin Kim}
\affiliation{Department of Physics, Sungkyunkwan University, Suwon 440-746, Korea}
\author{Kyeong Tae Kang}
\affiliation{Department of Physics, Sungkyunkwan University, Suwon 440-746, Korea}
\author{Jung Hoon Han}
\email[Electronic address:$~~$]{hanjh@skku.edu}
\affiliation{Department of Physics, Sungkyunkwan University, Suwon 440-746, Korea}

\begin{abstract} This note discusses several technical details not covered in the main text of the paper.
\end{abstract}
\maketitle

\section{Derivation of quasi-exact diagonalization}
In this section we derive the linear and the cubic orbital Rashba effects based on the tight-binding $t_{2g}$-band Hamiltonian.
The surface Hamiltonian without spin-orbit interaction is

\begin{align}\label{h1r}
&{\cal H} = \sum_{\v k, \sigma} \v C^\dag_{\v k, \sigma} {\cal H}_0 \v C_{\v k, \sigma}, \nn
&{\cal H}_0 = - \left(\begin{array}{ccc}
    2t(c_x \!+\!c_y)\!-\!\delta & -2i\gamma s_x  & -2i\gamma s_y \\
    2i\gamma s_x & 2tc_y\!+\!2t'c_x & 0 \\
    2i\gamma s_y  & 0 & 2tc_x\!+\!2t'c_y\\
  \end{array}\right) , \nn
\end{align}
where $\v C_{\v k,\sigma}=(c_{\v k,xy,\sigma}, c_{\v k,yz,\sigma} , c_{\v k,zx,\sigma})^T$ of spin $\sigma$.
Here $t$ and $t'$ represent $\sigma$-, $\pi$-bonding parameters, respectively, and the on-site energy difference $\delta$ is introduced due to the lack of electron hopping along the $z$-direction and the asymmetric crystal field.
Momentum dependencies are abbreviated as $c_{x(y)}=\cos k_{x(y)}$, $s_{x(y)}=\sin k_{x(y)}$, where lattice constant is taken to be unity.
Terms linear in $\gamma$ describe the interorbital hoppings due to the interface asymmetry, which can be cast as $\hat{z}\cdot(\mathbf{k} \times \mathbf{L})$ around the $\Gamma$ point~\cite{chiral-OAM}.
For simplicity of the ensuing calculation we replace the matrix elements in the Hamiltonian as

\begin{align}
    &e_1=-2t(c_x+c_y)+\delta,\nn
    &e_2=-2t c_y-2t' c_x,\nn
    &e_3=-2t c_x-2t' c_y,\nonumber
\end{align}
and then by taking out the energy $e_1$, the Hamiltonian reduces to

\begin{align}\label{reduceH}
    {\cal H}_0 = \left(\begin{array}{ccc}
    0                  & 2i\gamma s_x  & 2i\gamma s_y \\
    -2i\gamma s_x  & e_2 - e_1         & 0                \\
    -2i\gamma s_y  & 0                 & e_3 - e_1
    \end{array}\right),
\end{align}
where the appearance of $\v C_{\v k,\sigma}$ and $\v C^\dag_{\v k,\sigma}$ is implicit now.
It is useful to re-write the matrix elements in what we call the chiral basis, implemented through the unitary transformation

\renewcommand{\arraystretch}{1.7}
\begin{align}
U_1 = \bpm  \ds\frac{\Delta}{\sqrt{\Delta^2 + \gamma^2 s^2 }}     & \ds\frac{i \gamma s}{\sqrt{\Delta^2 + \gamma^2  s^2 } }       & 0 \\
\ds\frac{i \gamma s \cos \theta}{\sqrt{\Delta^2 + \gamma^2 s^2 }} & \ds\frac{ \Delta \cos \theta}{\sqrt{\Delta^2 +\gamma^2 s^2 }} & \sin \theta\\
\ds\frac{i \gamma s \sin \theta}{\sqrt{\Delta^2 + \gamma^2 s^2 }} & \ds\frac{\Delta \sin \theta}{\sqrt{\Delta^2 + \gamma^2 s^2 }} & -\cos \theta \epm  . \label{eq:U1}
\end{align}
The energy gap at the $\Gamma$ point between
$d_{xy}$-orbital-derived lowest-energy band and the rest is written
$\Delta=t-t'-\delta/2$, serving as a large energy denominator in the
perturbative scheme. The angle $\theta$ in the above is derived from

\begin{align}
\left(\cos\theta ,\sin\theta \right)=\left(s_x , s_y \right)/s, ~~ s= \sqrt{(s_x )^2 \!+ \! (s_y )^2 } . \label{eq:3}
\end{align}
In the new basis we obtain the matrix
\renewcommand{\arraystretch}{2.0}
\begin{widetext}
\begin{align}\label{eq:H-U1}
    {\cal H}_0^{(U_1) } = \left( \begin{array}{ccc}
    \ds\frac{\gamma^2 s^2 \left(e_p-8\Delta+e_m\cos2\theta \right)}{2\Delta^2}
    & \ds\frac{ i \gamma s \left(4\Delta-e_p-e_m\cos2\theta \right)}{2\Delta}  &  -\ds\frac{ie_m \gamma s \sin2\theta}{2\Delta} \\
    -\ds\frac{i \gamma s \left(4\Delta-e_p-e_m\cos2\theta \right)}{2\Delta}  & \ds\frac{e_p}{2} \left(1-\frac{\ds \gamma^2 s^2}{\Delta^2}\right) + \frac{\ds 8 \gamma^2 s^2 + e_m\Delta\cos2\theta}{2\Delta}
    &  \ds\frac{e_m \sin2\theta}{2}  \\
    \ds\frac{ie_m \gamma s \sin2\theta}{2\Delta}   &  \ds\frac{e_m \sin2\theta}{2}
     &  \ds\frac{e_p-e_m\cos2\theta}{2}
       \end{array} \right),
\end{align}
\end{widetext}
where
\begin{align}
    &e_p=e_2 + e_3 - 2e_1=(2t-2t')(c_x+c_y)-2\delta, \nn
    &e_m=e_2 - e_3=(2t-2t')(c_x-c_y).\nonumber
\end{align}
Terms having third or higher powers of $\gamma$ are neglected throughout the paper.
Besides, terms of order $k^4$ or higher have been dropped here.
The advantage of writing the Hamiltonian in the chiral basis is that the matrix elements connecting the first state to the rest are of cubic order in the momentum $\v k$.
Meanwhile, the second and the third states are nearly degenerate around the $\Gamma$ point, with off-diagonal elements of order $k^2$.
Pulling out a term $(2 e_p \Delta^2 + \alpha^2 \gamma^2 (8\Delta-e_p))/4\Delta^2$ from the diagonal elements, we are left with the matrix elements

\begin{widetext}
\begin{align}\label{eq:H-U1-1}
    \left(\begin{array}{ccc}
     \ds\frac{-2e_p\Delta^2+\gamma^2 s^2 \left(-24\Delta+3e_p+2e_m\cos2\theta \right)}{4\Delta^2}   & \ds\frac{i \gamma s \left(4\Delta-e_p-e_m\cos2\theta \right)}{2\Delta}  & -\ds\frac{i e_m \gamma s \sin2\theta }{2\Delta}  \\
      -\ds\frac{i \gamma s \left(4\Delta-e_p-e_m\cos2\theta \right)}{2\Delta}  & \sqrt{a^2+b^2}\cos \varphi &  \sqrt{a^2+b^2}\sin \varphi \\
      \ds\frac{i e_m \gamma s \sin2\theta }{2\Delta}  & \sqrt{a^2+b^2}\sin \varphi  &  -\sqrt{a^2+b^2}\cos \varphi \end{array}
    \right),\nn
\end{align}
\end{widetext}
where
\renewcommand{\arraystretch}{1.0}
\begin{align}\label{eq:phi-theta}
    &a=\frac{\gamma^2 s^2\left(8\Delta-e_p\right)+2e_m\Delta^2\cos2\theta}{4\Delta^2},\nn
    &b=\frac{e_m}{2}\sin2\theta ,\nn
    &\cos \varphi=\frac{a}{\sqrt{a^2+b^2}},~~~~\sin \varphi=\frac{b}{\sqrt{a^2+b^2}}.
\end{align}
Given the structure of the new matrix, we are prompted to adopt the second unitary rotation that diagonalizes the lower 2$\times$2 block of the Hamiltonian.
The second unitary matrix

\begin{align}\label{eq:V1}
V_1=\left(
    \begin{array}{ccc}
         1 &    0     & 0\\
         0 & \cos \varphi/2 & \sin \varphi/2 \\
         0 & \sin \varphi/2 & -\cos \varphi/2
    \end{array}\right)
\end{align}
renders the Hamiltonian

\begin{align}\label{eq:afterU1V1}
    {\cal H}_0^{(U_1 V_1 )}= \left( \begin{array}{ccc}
    E^{(1)}         & i \gamma S_x   &  i \gamma S_y \\
     -i \gamma S_x  &  E^{(-1)}      &  0  \\
     -i \gamma S_y  &  0             &  E^{(0)}
        \end{array} \right),
\end{align}
where

\begin{align}
    &E^{(1)} = \ds\frac{\gamma^2 s^2 \left(e_p-8\Delta+e_m\cos2\theta \right)}{2\Delta^2}, \nn
    &E^{(-1)} = \ds\frac{e_p}{2}\left(1-\frac{\gamma^2 s^2}{2\Delta^2}\right) + 2\frac{\gamma^2 s^2}{\Delta} + \xi, \nn
    &E^{(0)} = \ds\frac{e_p}{2}\left(1-\frac{\gamma^2 s^2}{2\Delta^2}\right) + 2\frac{\gamma^2 s^2}{\Delta} - \xi, \nn
    &S_x = \frac{s}{2\Delta}\left((4\Delta-e_p)\cos\frac{\varphi}{2} - e_m\cos\left[\frac{\varphi}{2} - 2\theta \right]\right), \nn
    &S_y = \frac{s}{2\Delta}\left((4\Delta-e_p)\sin\frac{\varphi}{2} - e_m\sin\left[\frac{\varphi}{2} - 2\theta \right]\right),\nn
    &\xi = \sqrt{\frac{e^2_m}{4} + \frac{e_m \gamma^2 s^2}{4\Delta^2}\left(8\Delta - e_p\right)\cos2\theta}. \nonumber
\end{align}
Here, $U_1 V_1 = {\cal U}_1$ in the main text. Choice of symbols $(S_x , S_y)$ reflect the analogous structure of ${\cal H}_0^{(U_1 V_1 )}$ to the original matrix ${\cal H}_0$. While $(s_x , s_y)$ were first order in $\v k$, the new $(S_x , S_y)$ are third-order. Motivated by the resemblance, we carry out a second set of unitary rotations, $U_2$ and $V_2$ ($U_2 V_2 = {\cal U}_2$ in the main text), serving the purposes of second chiral basis rotation similar to $U_1$ and diagonalizing the lower 2$\times$2 block, respectively.

Second chiral basis rotation matrix is

\renewcommand{\arraystretch}{1.9}
\begin{align}\label{eq:U2}
 U_2 =  \left(\begin{array}{ccc}
        \ds\frac{2 \Delta}{\sqrt{4 \Delta^2 + \gamma^2  \mathcal{S}^2}} & \ds\frac{i\gamma \mathcal{S}}{\sqrt{4 \Delta^2 + \gamma^2  \mathcal{S}^2}} & 0 \\
        \ds\frac{i \gamma \mathcal{S} \cos\Theta }{\sqrt{4 \Delta^2 + \gamma^2  \mathcal{S}^2}} & \ds\frac{2 \Delta \cos\Theta }{\sqrt{4 \Delta^2 + \gamma^2  \mathcal{S}^2}} & \ds\sin\Theta   \\
        \ds\frac{i  \gamma \mathcal{S} \sin\Theta }{\sqrt{4 \Delta^2 +   \gamma^2  \mathcal{S}^2}} & \ds\frac{2 \Delta \sin\Theta }{\sqrt{4 \Delta^2 + \gamma^2  \mathcal{S}^2}} &
        -\ds\cos\Theta
    \end{array}\right),
\end{align}
where $ \mathcal{S}=\sqrt{ S^2_x+ S^2_y}$ and $\left(\cos\Theta ,\sin\Theta \right)=\left( S_x, S_y\right)/ \mathcal{S}$. Matrix elements in the new basis read

\renewcommand{\arraystretch}{1.5}
\begin{align}\label{eq:U2-H}
    &{\cal H}_0^{(U_1 V_1 U_2)} =
        \left(\begin{array}{ccc}
        \mathcal{H}_{11}  & \mathcal{H}_{12}  & -i \ds\frac{\gamma \mathcal{S}}{2 \Delta} \xi \sin 2\Theta \\
        \mathcal{H}_{21}  & \mathcal{H}_{22}  &  \xi \sin 2\Theta  \\
        i \ds\frac{\gamma \mathcal{S}}{2 \Delta} \xi \sin 2\Theta & \xi \sin 2\Theta  &  \mathcal{H}_{33}
        \end{array}\right),\nn
\end{align}
where

\begin{align}
    &\mathcal{H}_{11}=\ds\frac{\gamma^2 s^2}{2\Delta^2}\left(e_p-8\Delta+e_m\cos2\theta\right),\nn
    &\mathcal{H}_{22}=\ds\frac{e_p}{2}\left(1-\frac{\gamma^2 s^2}{2\Delta^2}\right) + 2\frac{\gamma^2 s^2}{\Delta} + \xi\cos2\Theta,\nn
    &\mathcal{H}_{33}=\ds\frac{e_p}{2}\left(1-\frac{\gamma^2 s^2}{2\Delta^2}\right) + 2\frac{\gamma^2 s^2}{\Delta} - \xi\cos2\Theta,\nn
    &\mathcal{H}_{21}=\mathcal{H}_{21}^* =\ds\frac{i\gamma \mathcal{S}}{4\Delta}\left( e_p - 4\Delta + 2\xi\cos2\Theta \right).\nonumber
\end{align}
Terms of order $k^6$ or higher have been dropped.
Subsequent unitary matrix

\renewcommand{\arraystretch}{1.0}
\begin{align}
    V_2= \left(\begin{array}{ccc}
    1   &   0           &   0                \\
    0   &   \cos\Theta    &   \sin\Theta  \\
    0   &   \sin\Theta    &   -\cos\Theta \\
    \end{array}\right)
\end{align}
leads to the Hamiltonian

\begin{align}\label{eq:V2-H}
   &{\cal H}_0^{(U_1 V_1 U_2 V_2)} =  \left(\begin{array}{ccc}
    E^{(1)}   & \mathcal{H'}_{12}              & \mathcal{H'}_{13}  \\
    \mathcal{H'}_{21}              & E^{(-1)}   & 0 \\
    \mathcal{H'}_{31}              &  0                             & E^{(0)}
    \end{array}\right),
\end{align}

where
\begin{align}
    &\mathcal{H'}_{21} = \mathcal{H'}_{12}^* =i \frac{\gamma \mathcal{S}}{4 \Delta}\left(6\frac{\gamma^2 s^2}{\Delta} + 2\xi + e_p - 4\Delta \right) \cos\Theta, \nn
    &\mathcal{H'}_{31} = \mathcal{H'}_{13}^* =i \frac{\gamma \mathcal{S}}{4 \Delta}\left(6\frac{\gamma^2 s^2}{\Delta} - 2\xi + e_p - 4\Delta \right) \sin\Theta. \nonumber
\end{align}
Diagonal elements are the same as in Eq.~(\ref{eq:afterU1V1}).
The remaining off-diagonal elements are proportional to $\gamma \mathcal{S}$ which scales as $k^3$.
In addition, terms inside the parenthesis of $\mathcal{H'}_{21}$ and $\mathcal{H'}_{31}$ also give $k^2$ dependence, since $\xi \sim k^2$ and $(e_p-4\Delta) \sim k^2$.
It is shown that the off-diagonal elements are of $k^5$ power, beyond the order of approximation we are interested in.
They can be thus neglected, leaving us with an ``exactly diagonalized" Hamiltonian and a set of ``exact" eigenstates which one can use to compute various averages.
Recall that dropping the off-diagonal terms from the above Hamiltonian is the only approximation we are using to diagonalize the original matrix. To the extent that we are concerned with effects of order $k^3$ at the most, the method employed here (a series of unitary transformations) is exact.

\vspace{10mm}

\section{Orbital angular momentum}

Orbital angular momentum (OAM) averages of each band are given by the expectation values of $\v L = (\sum_i \v L_i )/N$, for the atomic OAM operator $\v L_i$ in the $t_{2g}$-orbital space at site $i$, and $N$ denoting the number of atomic sites~\cite{chiral-OAM}.
Based on the diagonalization scheme that led to Eq. (\ref{eq:V2-H}) we can work out the OAM averages

\begin{align}\label{eq:OAM-general}
L_+^{(1)} &= -2 i \frac{\gamma s}{\Delta} e^{i\theta} - 2i \frac{\gamma \mathcal{S}}{\Delta} e^{i (\theta + \Theta - \varphi/2) }, \nn
L_+^{(0)} &=  i \frac{\gamma s}{\Delta} \bigl( e^{i \theta } - e^{i (\theta - \varphi)} \bigr)  - 2\frac{\gamma \mathcal{S}}{\Delta} \sin\Theta \;e^{i (\theta -\varphi /2) } , \nn
L_+^{(-1)} &= i \frac{\gamma s}{\Delta} \bigl( e^{i \theta } + e^{i (\theta - \varphi)} \bigr) + 2 i \frac{\gamma \mathcal{S}}{\Delta} \cos \Theta \;e^{i(\theta - \varphi/2 ) } ,
\end{align}
where  $L_+^{(m)}$ denotes $\langle \psi^{(m)} | (L_x \! + \! i L_y ) |\psi^{(m)} \rangle$, an average for the quasi-exact eigenstate $|\psi^{(m)} \rangle$ with energy $E^{(m)}$.
Terms of order $k^5$ or higher have been dropped.
Average of $L_z$ is strictly zero for all three bands.
Sum of OAM averages is identically zero.
The power-law dependencies are $\gamma s / \Delta\sim k^1$ and $\gamma \mathcal{S}/\Delta \sim k^3$.

One sees upon examination of the general result (\ref{eq:OAM-general}) that the OAM averages depend on $\varphi$ and $\theta$ (angle in the
momentum space), and $\varphi$ in turn depends on $\theta$ through the relation (\ref{eq:phi-theta}).
Actually Eq. (\ref{eq:phi-theta}) allows two limiting cases near the $\Gamma$ point for analysis, $\gamma\gg \Delta$ and $\gamma\ll \Delta$.
At first, we expand $s$'s around the $\Gamma$ point as

\begin{align}
s_x & \sim   k_x  - k_x^3 /6  , \nn
s_y & \sim   k_y - k_y^3 /6 , \nn
s &\sim  k -  k^3 ( \cos^4 \theta_k + \sin^4 \theta_k ) /6, \nonumber
\end{align}
hence
\begin{align}
\cos \theta &\sim \cos \theta_k + \frac{k^2}{6} \left(\cos^5 \theta_k + \cos \theta_k \sin^4 \theta_k - \cos^3 \theta_k \right),  \nn
\sin \theta &\sim \sin \theta_k + \frac{k^2}{6} \left(\cos^4 \theta_k \sin \theta_k + \sin^5 \theta_k - \sin^3 \theta_k \right), \nonumber
\end{align}
where
$\left(\cos\theta_k,\sin\theta_k\right)=\left(k_x,k_y\right)/k$.
There is a $k^3$ dependence coming out of $s$, which will be taken into account in the OAM averages.

In the $\gamma\gg \Delta$ limit, OAM averages simplify greatly as

\begin{align}\label{eq:OAM-large-gamma}
    &L_+^{(1)} \approx -2 i \frac{\gamma}{\Delta} \mathcal{M} k_+ + i \frac{\gamma}{\Delta} \left[\frac{1}{12} - \frac{t - t'}{4\Delta}\right] k_-^3 ,\nn
    &L_+^{(-1)} \approx \; 2 i \frac{\gamma}{\Delta} \mathcal{M} k_+ - i \frac{\gamma}{\Delta} \left[\frac{1}{12} - \frac{t - t'}{8\Delta}\right] k_-^3, \nn
    &L_+^{(0)} \approx \;i \frac{\gamma}{\Delta} \frac{t-t'}{8\Delta}k_-^3 , \nn
    &\mathcal{M} = 1 - \frac{\Delta - 3(t - t')}{8 \Delta}k^2,
\end{align}
where $k_\pm = k_x \pm i k_y$. Now one can see each superscript in $|\psi^{(m)} \rangle$ reflects not only the energy band index but also the orbital chirality. $\mathcal{M}$ is simplified to be $1$ in the main text because $k^2$ term in $\mathcal{M}$ is less important than $1$ near the $\Gamma$ point.

In the other limit $\gamma\ll \Delta$, OAM averages become

\begin{align}\label{eq:OAM-small-gamma}
&L_+^{(-1)} ~(L_+^{(0)}) \approx \left\{
                      \begin{array}{ll}
                        P_x ~(P_y), & \hbox{$|k_x|>|k_y|$,} \\
                        P_y ~(P_x), & \hbox{$|k_x|<|k_y|$,} \\
                        P_x+P_y ~(0), & \hbox{$|k_x|=|k_y|$,} \\
                      \end{array}
                    \right. \nn
& P_x = i \frac{\gamma}{\Delta} \mathcal{M} (k_+ - k_-) + \frac{i}{2} \frac{\gamma}{\Delta} \left[\frac{1}{12} - \frac{t-t'}{4\Delta}\right](k_+^3 - k_-^3 ) ,\nn
& P_y = i \frac{\gamma}{\Delta} \mathcal{M} (k_+ + k_-) - \frac{i}{2} \frac{\gamma}{\Delta} \left[\frac{1}{12} - \frac{t-t'}{4\Delta}\right](k_+^3 + k_-^3 ) .
\end{align}
$L_+^{(1)}$ takes the same form as in Eq.~(\ref{eq:OAM-large-gamma}).

\section{Numerical method}

\begin{figure*}[htbp]
\includegraphics[width=0.98\textwidth]{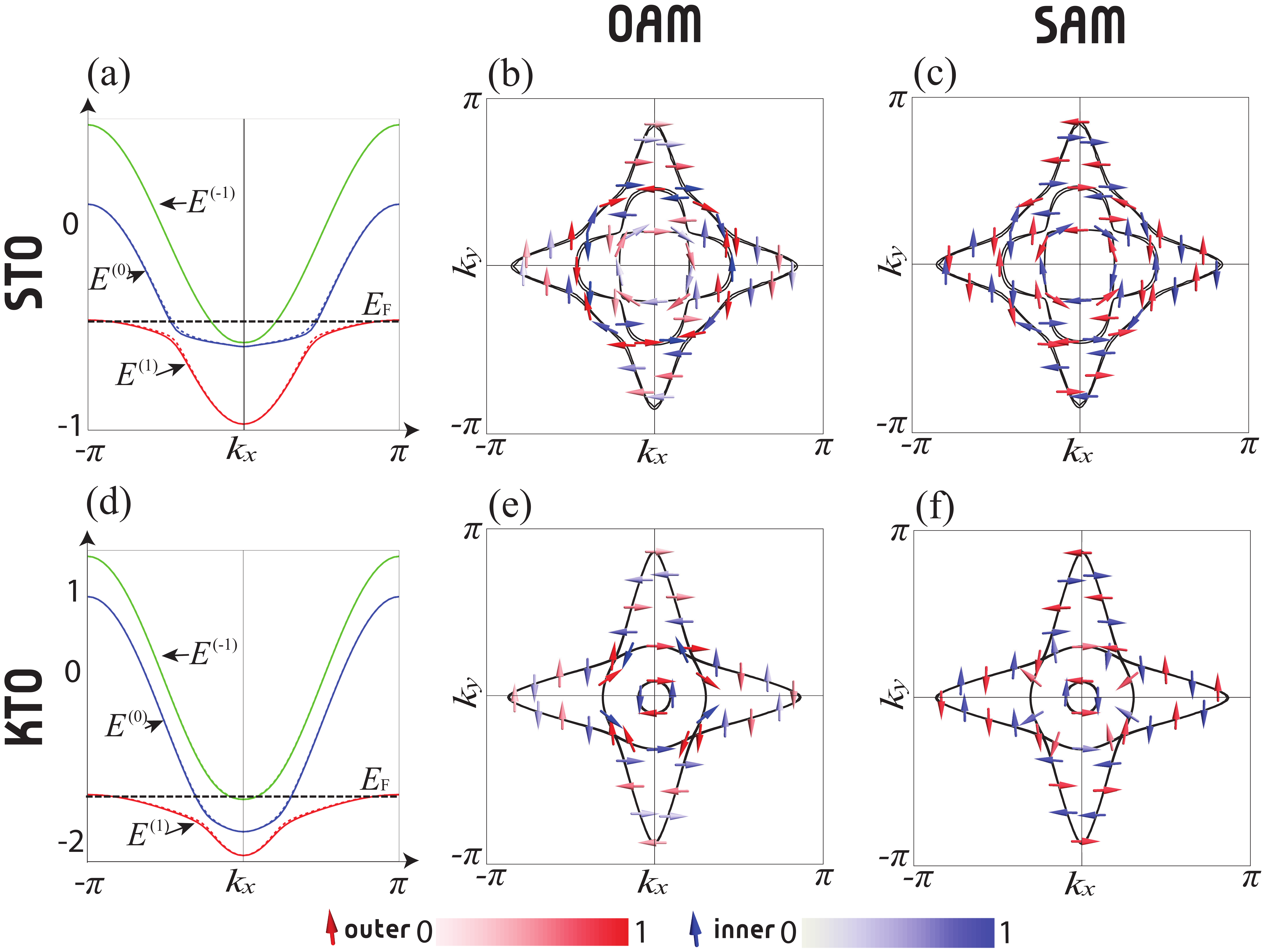}
\caption{(color online) (a)-(f) Band, OAM and SAM structures of (a)-(c) STO and (d)-(f) KTO.
Fermi levels are determined pertinent to the surface electronic structures from previous experiments~\cite{king-texture,KTOprb}.
Band and OAM structures are discussed in the main text.
Inclusion of SOI gives rise to the similar SAM aspects in STO and KTO in the sense that SAM is fully polarized in opposite directions in each Rashba-split states, either parallel or anti-parallel to the underlying OAM.
Note, however, that in KTO near the points $E^{(1)}$ and $E^{(0)}$ surface states are close, OAM and SAM directions are not parallel nor anti-parallel.
}
\label{fig:TB-AM}
\end{figure*}

Numerical diagonalization is performed to investigate the OAM and spin angular momentum (SAM) structures of SrTiO$_3$ (STO) and KTaO$_3$ (KTO) based on the tight-binding model.
Multi-orbital Hamiltonian (\ref{h1r}) together with spin-orbit interaction, ${\cal H}_{\rm so} = \lambda_{\rm so} \v L \cdot \bm \sigma$, where parameters given by the Table~I from the main text, is fully diagonalized in STO and KTO.
OAM and SAM structures are summarized in Fig.~\ref{fig:TB-AM}.

